\documentclass[10pt,journal,compsoc]{IEEEtran}
\usepackage[utf8]{inputenc}
\usepackage{amsmath,amsfonts}
\usepackage{array}
\usepackage{url}
\usepackage{graphicx}
\usepackage{cite}
\usepackage{framed}
\usepackage{tikz}
\usetikzlibrary{shapes.geometric}
\usepackage{booktabs}
\usepackage{multirow}
\usepackage{hyperref}
\usepackage{enumitem}
\usepackage[table]{xcolor}
\usepackage[nolist]{acronym}

\definecolor{drawio_yellow}{HTML}{FFE000}
\definecolor{drawio_red}{HTML}{E51400}
\definecolor{drawio_blue}{HTML}{0050EF}
\definecolor{drawio_green}{HTML}{008A00}
\definecolor{drawio_styles_red_border}{HTML}{B95450}
\definecolor{drawio_styles_red_fill}{HTML}{F8CECC}
\definecolor{drawio_styles_yellow_fill}{HTML}{FFF2CC}
\definecolor{drawio_styles_yellow_border}{HTML}{D6B656}
\definecolor{drawio_styles_blue_fill}{HTML}{DAE8FC}
\definecolor{drawio_styles_blue_border}{HTML}{6C8EBF}

\definecolor[named]{ACMBlue}{cmyk}{1,0.1,0,0.1}
\definecolor[named]{ACMYellow}{cmyk}{0,0.16,1,0}
\definecolor[named]{ACMOrange}{cmyk}{0,0.42,1,0.01}
\definecolor[named]{ACMRed}{cmyk}{0,0.90,0.86,0}
\definecolor[named]{ACMLightBlue}{cmyk}{0.49,0.01,0,0}
\definecolor[named]{ACMGreen}{cmyk}{0.20,0,1,0.19}
\definecolor[named]{ACMPurple}{cmyk}{0.55,1,0,0.15}
\definecolor[named]{ACMDarkBlue}{cmyk}{1,0.58,0,0.21}
\hypersetup{
    colorlinks=true,
    citecolor=ACMPurple,    
    linkcolor=ACMBlue,      
    urlcolor=ACMDarkBlue    
}

\newcommand{\coloredcircle}[3]{%
    \tikz[baseline=(char.base)]{
        \node[shape=circle,inner sep=1pt,fill=#2,text=#3] (char) {#1};
    }
}
\newcommand{\coloredhexagon}[3]{%
   \tikz[baseline=(char.south)]{
       \node[shape=regular polygon,regular polygon sides=6,inner sep=2pt,fill=#2,draw=#3,text=#3] (char) {#1};
   }
}

\begin{acronym}[eng]
    \acro{ADR}{Architecture Decision Record}
    \acro{AI}{Artificial Intelligence}
    \acro{CNN}{Convolutional Neural Network}
    \acrodefplural{CNN}[CNNs]{Convolutional Neural Networks}
    \acro{CRHF}{Collision-Resistant Hash Function}
    \acrodefplural{CRHF}[CRHFs]{Collision-Resistant Hash Functions}
    \acro{DLRM}{Deep Learning Recommendation Model}
    \acro{DNN}{Deep Neural Network}
    \acrodefplural{DNN}[DNNs]{Deep Neural Networks}
    \acro{EPC}{Enclave Page Cache}
    \acro{LLM}{Large Language Model}
    \acrodefplural{LLM}[LLMs]{Large Language Models}
    \acro{ML}{Machine Learning}
    \acro{MLaaS}{Machine Learning as a Service}
    \acro{MLOps}{Machine Learning Operations}
    \acro{MLP}{Multilayer Perceptron}
    \acro{MPC}{Multi-Party Computation}

    \acro{ONNX}{Open Neural Network Exchange}
    \acro{ReLU}{Rectified Linear Unit}
    \acro{SBOM}{Software Bill of Materials}
    \acro{SGD}{Stochastic Gradient Descent}
    \acro{SGX}{Software Guard Extensions}
    \acro{SMPC}{Secure Multi-Party Computation}
    \acro{SNARK}{Succinct Non-interactive Argument of Knowledge}
    \acrodefplural{SNARK}[SNARKs]{Succinct Non-interactive Arguments of Knowledge}
    \acro{STARK}{Scalable Transparent Argument of Knowledge}
    \acro{TDSP}{Team Data Science Process}
    \acro{TEE}{Trusted Execution Environment}
    \acrodefplural{TEE}[TEEs]{Trusted Execution Environments}
    \acro{VC}{Verifiable Computation}
    \acro{VV}[V\&V]{Verification and Validation}
    \acro{ZKML}{Zero-Knowledge Machine Learning}
    \acro{ZKP}{Zero-Knowledge Proof}
    \acrodefplural{ZKP}[ZKPs]{Zero-Knowledge Proofs}

\end{acronym}

\usepackage[utf8]{inputenc}
\begin{document}

\title{``Show Me You Comply\ldots Without Showing Me Anything'': Zero-Knowledge Software Auditing for AI-Enabled Systems}

\author{Filippo Scaramuzza,~Renato Cordeiro Ferreira,~Giovanni Quattrocchi,~Damian Andrew Tamburri,~and~Willem-Jan van den Heuvel%
\thanks{F. Scaramuzza is with Tilburg University, Tilburg, Netherlands, and Eindhoven University of Technology, Eindhoven, Netherlands. E-mail: f.scaramuzza@uvt.nl}%
\thanks{R.~C. Ferreira is with Tilburg University, Tilburg, Netherlands; Eindhoven University of Technology, Eindhoven, Netherlands; and University of São Paulo, São Paulo, Brazil.}%
\thanks{G. Quattrocchi is with Politecnico di Milano, Milano, Italy.}%
\thanks{D.~A. Tamburri is with University of Sannio, Benevento, Italy, and Eindhoven University of Technology, Eindhoven, Netherlands.}%
\thanks{W.-J. van den Heuvel is with Tilburg University, Tilburg, Netherlands, and Eindhoven University of Technology, Eindhoven, Netherlands.}%
}

\markboth{IEEE Transactions on Software Engineering}%
{Scaramuzza \MakeLowercase{\textit{et al.}}: Zero-Knowledge Software Auditing for AI-Enabled Systems}

\maketitle

\begin{abstract}
Classical software verification and validation techniques, such as procedural audits, formal methods, or model documentation, are the traditional mechanisms used to achieve the verifiable accountability now required by regulations like the EU AI Act. These methods are either expensive or heavily manual, and ill-suited for the opaque, ``black box'' nature of most \ac{AI} models. A conflict arises: high auditability and verifiability are required by law, but such transparency conflicts with the need to protect the assets being audited (e.g., confidential data and proprietary models). This paper introduces ZKMLOps, an \ac{MLOps} verification framework that operationalizes \acp{ZKP} within Machine-Learning Operations lifecycles; a \ac{ZKP} allows a prover to convince a verifier that a statement is true without revealing any information about the statement itself. By integrating \acp{ZKP} with established software engineering patterns, ZKMLOps provides a modular and repeatable process for generating verifiable cryptographic \emph{evidence}---proofs of well-defined computational statements about the audited model and its inputs---that auditors can use as input to a regulatory compliance determination. We evaluate the framework along two dimensions. First, framework viability: orchestration overhead is bounded and stable across architecturally heterogeneous ZKP backends and models of increasing size. Second, cost-versus-assurance trade-offs: the audit-on-demand setting is the regime in which full zero-knowledge auditing is the appropriate tool, where it provides confidentiality and integrity guarantees that lighter-weight alternatives cannot match.
\end{abstract}


\section{Introduction}
\label{sec:introduction}

\IEEEPARstart{A}{I-enabled} systems are increasingly safety-, ethics-, and/or money-critical~\cite{Amodei2018,Lipton2018}, thus legislators worldwide have begun to demand \emph{verifiable accountability}. For instance, the EU Artificial Intelligence Act\footnote{Regulation~(EU)~2024/1689, enforced from~June~2024} classifies many industrial AI deployments as \emph{high-risk} and requires evidence of compliance with strict transparency, robustness, and human-oversight obligations~\cite{EUArtificialIntelligence}. 

In conventional software engineering, auditors rely on a spectrum of \ac{VV} artifacts, ranging from manual process audits and test suites to formal model-checking~\cite{Clarke2018}.
Such artifacts are, however, only effective when the implementation logic is explicit and can be modelled deterministically.
Yet modern AI models are opaque ``black boxes'' whose behaviour emerges from millions (or even billions) of learned parameters~\cite{Rudin2019}.
Probing these parameters directly may leak intellectual property or personal data, and disclosing training data often conflicts with data-protection laws and contracts.
Organizations face a hard tension: regulators request deep transparency, but disclosing the very artifacts that enable transparency would compromise proprietary or privacy-sensitive assets~\cite{cooper_accountability_2022, cofone_strategic_2019}.
This gap has real-world consequences: the 2015 Volkswagen emissions scandal~\cite{jung2019volkswagen} showed how software can be programmed to behave differently under audit conditions than in production, a pattern of engineered deception equally possible in opaque \ac{ML} systems~\cite{mokander_auditing_2023}.

This paper introduces ZKMLOps, a framework that operationalizes \acp{ZKP}~\cite{zkmlSurvey2025} within the \ac{MLOps} lifecycle~\cite{kreuzberger_machine_2023}, enabling an organization to provide verifiable cryptographic evidence on specific computational properties of its systems, such as whether an approved model produced a given inference on approved input data, which auditors can then assess against regulatory requirements. A ZKP allows a prover to convince a verifier that a computational statement is true without revealing any information beyond the statement's validity, which fits the auditing dilemma. However, the availability and active research of these primitives~\cite{Groth2016,Gabizon2020,zkmlSurvey2025} does not address the practical challenge of their systematic application. 

This paper makes three contributions. First, we propose ZKMLOps, a framework that integrates ZKP-based verification into the MLOps lifecycle and show its application in a use case scenario in financial risk auditing. Second, we provide an evaluation of the framework's practicality. Our benchmarks show that ZKMLOps adds a stable orchestration overhead, which becomes negligible as circuit complexity increases. Third, we identify the specific regulatory and technical regimes where full \acp{ZKP} outperfomes alternative methods such as \acp{TEE} and sampling with \ac{VC}.

The remainder of this paper is organized as follows. Section~\ref{sec:background} introduces background on trustworthy AI, ML auditing, and zero-knowledge proofs. Section~\ref{sec:related_work} reviews related work and positions ZKMLOps within the existing landscape. Section~\ref{sec:methodology} presents the research methods and central research question. Section~\ref{sec:framework} describes the proposed framework. Section~\ref{sec:use-case} applies it to a regulatory compliance use case. Section~\ref{sec:evaluation} reports the empirical evaluation. Section~\ref{sec:discussion} discusses the findings. Section~\ref{sec:implications} draws implications for practice. Section~\ref{sec:threats} addresses threats to validity, and Section~\ref{sec:conclusions} concludes.

\section{Background}
\label{sec:background}

\subsection{Trustworthy AI}

According to the European Union~\cite{smuha2019eu}, an ML-enabled system should meet four ethical principles: (i) respect for human autonomy, (ii) prevention of harm, (iii) fairness, and (iv) explicability. Building on these, Liu et al.~\cite{liu_trustworthy_2022} survey recent research and identify six dimensions of trustworthy AI: (i) \emph{Safety and Robustness}, i.e., resilience to small input perturbations and the ability to make secure decisions; (ii) \emph{Non-discrimination \& Fairness}, i.e., avoidance of unfair bias toward groups or individuals; (iii) \emph{Explainability}, i.e., the ability to be explained to stakeholders; (iv) \emph{Privacy}, i.e., not leaking private information; (v) \emph{Accountability}, i.e., being assessable by a third party and assigning responsibility for an AI failure; (vi) \emph{Environmental Well-Being}, i.e., being sustainable and environmentally friendly.

This work focuses on accountability, since it covers regulatory requirements that bind the other dimensions. As Li et al.~\cite{li_trustworthy_2023} note, accountability runs through the entire lifecycle of an ML-enabled system and requires its stakeholders to justify product design, software architecture, implementation, and operation as aligned with human values.

From accountability are also derived the concepts of auditability and traceability, which require an ML-enabled system to be reviewed and audited~\cite{leslie_understanding_2019}.

\subsection{Auditing of ML-Enabled Systems}

The auditing of an ML-enabled system (AI auditing for simplicity) can be defined along two axes: functionally and methodologically~\cite{mokander_auditing_2023}. Functionally, AI auditing is a governance mechanism that can be managed by various actors. For instance, regulators may assess compliance with legal standards, technology providers can investigate how to mitigate technology-related risks, and stakeholders can make informed decisions about how they engage with specific companies. Methodologically, AI auditing is characterized by a structured process whereby an entity's past or present behavior is assessed for consistency with predefined standards or regulations.

M\"okander et al.~\cite{mokander_auditing_2023} identify two main drivers for AI auditing: it serves as (i) a mechanism to implement legislation and (ii) a governance mechanism for private corporations. The first is a top-down pressure, as more regulations are released, such as the EU AI Act~\cite{EUArtificialIntelligence} or Canada's Directive on Automated Decision-Making~\cite{secretariat_directive_2024}. The second is bottom-up: organizations that deploy \ac{AI} systems have incentives to maintain governance mechanisms; by undergoing independent audits, they can assess and improve their software development process and quality management system~\cite{vlok2003technology}.

Per definition, auditing requires a predefined baseline, which can consist of technical specifications, legal requirements, or voluntary ethics principles, giving rise to three categories~\cite{mokander_auditing_2023}: (i)~\emph{technical} approaches, which quantify properties like accuracy and robustness through \ac{VV} practices~\cite{brundage_toward_2020}; (ii)~\emph{legal} approaches, which assess regulatory compliance; and (iii)~\emph{ethical} approaches, which evaluate against voluntary principles such as those from IEEE~\cite{chatila2019ieee}.

In practice, these three types of approaches have loose boundaries. For example, to establish legal compliance, auditors typically rely on technical methods for gathering evidence about the properties and impact ML-enabled systems have~\cite{kim2017auditing}. We place our contribution in this last scenario, where the verification of specific properties throughout the MLOps verification lifecycle can serve as a mechanism for legal compliance assessment.

\subsection{Zero-Knowledge Proofs}
ZKPs provide a formal mechanism through which a \emph{prover} can convince a \emph{verifier} that a given statement is true, without revealing any information beyond the truth of the statement itself~\cite{goldreich_foundations_2001}.

At the core of modern ZKP systems is the transformation of computations into \emph{arithmetic circuits} over a finite field $\mathbb{F}_p$~\cite{weizmann_institute_of_science_knowledge_2019}. Any computable function can be rewritten as a sequence of additions and multiplications over $\mathbb{F}_p$, where $p$ is a large prime. Formally, the prover proves the existence of a secret witness $w$ satisfying:
$$
C(x, w) = y
$$
where $C$ is the circuit, $x$ the public input, and $y$ the public output; the verifier accepts if and only if $C(x,w)$ holds, without learning $w$.

The Fiat--Shamir heuristic~\cite{fiat_how_1987} transforms interactive ZKPs into non-interactive ones, yielding a self-contained proof any verifier can check independently.

Polynomial commitment schemes~\cite{kate_constant-size_2010} encode the computation trace into a committed polynomial; the verifier checks compliance by querying a few evaluations, reducing proof size and verification cost, achieving \emph{succinctness}.

A key challenge in applying ZKPs to domains such as ML is handling \emph{non-linear functions}, which are not naturally supported in arithmetic circuits. Neural networks, for example, often include non-linear activation functions like the Rectified Linear Unit ($ReLU(x) = \max(0, x)$)~\cite{arora_understanding_2018}.
To represent such operations in ZKP-friendly form, systems typically use \emph{lookup arguments}~\cite{kang_scaling_2022}. In a lookup argument, the prover shows that each non-linear operation maps an input to an output according to a precomputed table~$T$:
\[
\exists (x, y) \in T \quad \text{such that} \quad y = f(x)
\]
This allows incorporating non-polynomial logic into ZKPs while preserving succinctness and zero-knowledge.

\section{Related Work}
\label{sec:related_work}

\subsection{ZKP-based ML Verification}
\label{sec:related_zkml}

The literature has investigated the applications of \ac{ZKP} techniques to \ac{ML}, an area known as \ac{ZKML}. The objective is to use ZKPs in ML to prove that computations within an ML system, such as inference, training, or data preprocessing, have been carried out correctly without disclosing the underlying data, model parameters, or internal workings~\cite{xing2025zero, scaramuzza_engineering_2025}.
Scaramuzza et al.~\cite{scaramuzza_engineering_2025} first survey the existing literature on ZKP protocols and identify five properties (non-interactivity, transparent setup, standard representations, succinctness, and post-quantum security) that matter for their application in ML validation and verification pipelines. They then run a follow-up systematic survey analyzing ZKP-enhanced ML applications across an adaptation of the Microsoft Machine Learning Workflow model (Data \& Preprocessing, Training \& Offline Metrics, Inference, and Online Metrics), detailing verification objectives, ML models, and adopted protocols. This taxonomy of properties underpins the Property Matrix used throughout this paper; the survey itself does not provide a practical implementation or guidelines for operationalizing such systems.

The current literature places strong emphasis on inference verification, the most extensively explored subfield of ZKML. Examples include ZEN~\cite{feng_zen_2021}, vCNN~\cite{lee_vcnn_2024}, zkCNN~\cite{liu_zkcnn_2021}, and~\cite{sun_zkllm_2024}. More recent work covers training~\cite{abbaszadeh_zero-knowledge_2024} and online metrics verification~\cite{zhang_zero_2020}.

Waiwitlikhit et al.~\cite{waiwitlikhit_trustless_2024} propose \textsc{ZKAudit}, a zero-knowledge audit framework for trustless verification of model training and data properties without revealing model weights or training data. The framework supports real-world models (MobileNet~v2, \ac{DLRM}) and enables audits such as censorship detection, copyright verification, and counterfactual analysis.
ZKAudit is a \ac{ZKP} construction (a Halo2-based circuit family with prover-side optimisations for \ac{SGD} training and arbitrary-property auditing) which could, in principle, be plugged into ZKMLOps a workflow. 

\subsection{Adjacent Privacy-Preserving Auditing Approaches}
\label{sec:related_adjacent}

Guo et al.~\cite{guo_immaculate_2024} propose IMMACULATE, a framework combining randomized inference sampling with verifiable-computation footprints per sampled request, targeting high-throughput streaming inference. Each sampled inference carries a full ZK proof while the remainder of the stream is bound to a hash-chain root, so the amortised steady-state cost per request scales with the sampling rate~$\alpha$. The trust model differs from full ZK: the auditor accepts a statistical guarantee in exchange for orders-of-magnitude lower per-request cost.

Lycklama et al.~\cite{lycklama_arc_2024} introduce Arc, a system that generates concise auditing receipts for ML computations using cryptographic commitments and statistical arguments, without the full proof-generation cost of a zk-SNARK. The trust model is crypto-only (no hardware or non-collusion assumption), but argument size grows with the required statistical confidence rather than being constant in circuit depth. Arc sits in the same trust regime as full ZK, with speedups up to~$10^4\times$ relative to hash-based commitment approaches.

Pal et al.~\cite{pal_priveri_2024} propose Priveri, a sampling-based framework for privacy-preserving verifiable inference over large language models. Priveri runs the base inference under a \ac{SMPC} protocol among non-colluding parties and inserts sentinel inputs at random positions to detect deviations from the committed model; verification is probabilistic rather than cryptographically binding. Compared to full ZK, Priveri is substantially cheaper per token (22.1~s wall-clock on Llama-2-7B) but offers only soundness against a polynomial number of queries.

Toreini et al.~\cite{toreini_fairness_2024} verify fairness metrics (Demographic Parity, Equalized Odds) on a public board without revealing the model or its data: participants submit encrypted cryptograms, each backed by a ZK proof of well-formedness. The system scales to over~1\,000 participants and is the closest existing work to ZKMLOps's workflows. While the main contribution of FaaS is the fairness circuit and the cryptogram protocol, ZKMLOps could be the orchestration layer that makes circuits like this one deployable as part of a repeatable MLOps auditing process.

\section{Research Methods}
\label{sec:methodology}

\begin{figure}[b]
    \centering
    \includegraphics[width=.9\linewidth]{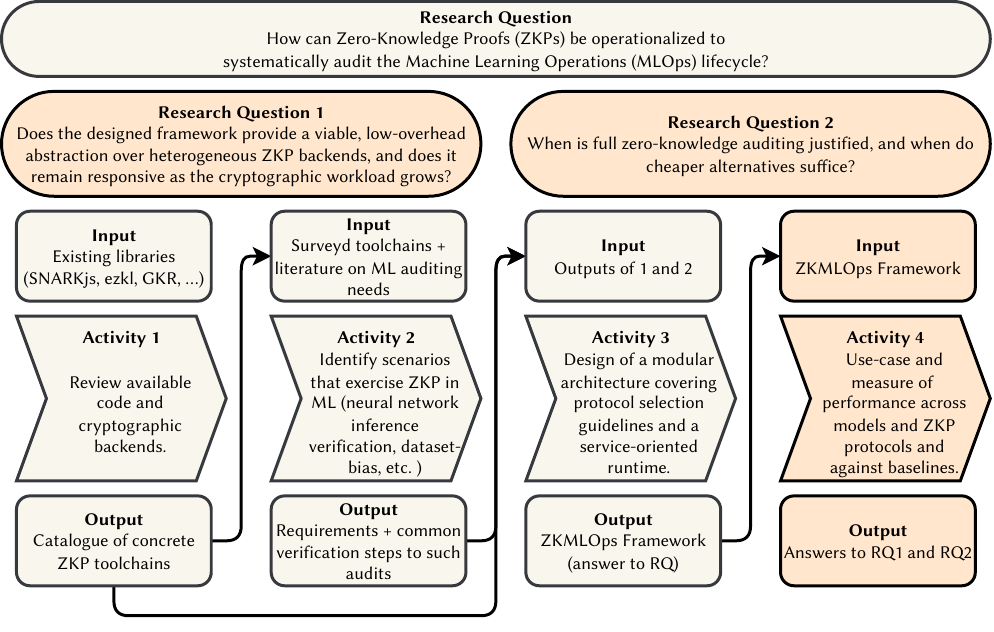}
    \caption{Activity Diagram of the four-step approach for this study.}
    \label{fig:methodology}
\end{figure}

Our approach is grounded on the following \emph{research question}:
\begin{description}
    \item[\textbf{RQ}] \textit{How can Zero-Knowledge Proofs (ZKPs) be operationalized to systematically audit the Machine Learning Operations (MLOps) lifecycle?}
\end{description}
Despite the maturation of ZKP protocols, there are not frameworks addressing their systematic integration into the repeatable engineering processes that MLOps teams require in production. Answering this question requires bridging cryptographic protocol research and the software engineering lifecycle. The main \textbf{RQ} is then further developed in two sub-questions:
\begin{description}
  \item[\textbf{RQ1}] Does the designed framework provide a viable, low-overhead abstraction over heterogeneous ZKP backends, and does it remain responsive as the cryptographic workload grows?
  \item[\textbf{RQ2}] When is full zero-knowledge auditing justified, and when do cheaper alternatives suffice?
\end{description}

\noindent The study follows four steps, depicted in Figure~\ref{fig:methodology}.

\textit{Step 1: Toolchain survey.} We reviewed existing ZKP protocol implementations (ezkl, snarkjs, Cairo/Stwo) and their supporting libraries, mapping each against the five protocol-selection criteria from Scaramuzza et al.~\cite{scaramuzza_engineering_2025}: non-interactivity, transparent setup, standard representation, succinctness, and post-quantum security. This grounded the framework design in concrete, deployable backends rather than theoretical constructions.

\textit{Step 2: Use case analysis.} We identified regulatory compliance in financial risk modeling as the primary audit scenario and derived the concrete requirements for the framework: what artifacts must be committed, what properties must be proven, and what evidence the auditor needs. This step produced the use case described in Section~\ref{sec:use-case} and drove the protocol-selection guidelines in Section~\ref{sec:guidelines}.

\textit{Step 3: Framework design.} We designed ZKMLOps using hexagonal architecture~\cite{martin_clean_2018}, the Orchestrated Saga pattern~\cite{richardson_microservices_2018}, and the State and Strategy patterns~\cite{gamma1995design} to decouple the core auditing logic from ZKP backend implementations. The artifact-binding mechanism and trust model (Section~\ref{sec:trust-model}) were specified as part of this step.

\textit{Step 4: Empirical evaluation.} We assessed the framework along two dimensions. First, we measured orchestration overhead across architecturally distinct backends (ezkl/Halo2 and snarkjs/Groth16) on tabular models and a MobileNet-v2 scaling sweep, to determine whether the abstraction layer imposes a bounded, backend-invariant cost. Second, we compared full ZK against lighter-weight alternatives (sampling-with-\ac{VC}, \ac{TEE}, \ac{MPC}) to map the cost-versus-assurance trade-off and identify the settings where each technique is appropriate.

\section{Proposed Framework}
\label{sec:framework}

\begin{figure}[!t]
    \centering
    \includegraphics[width=0.85\linewidth]{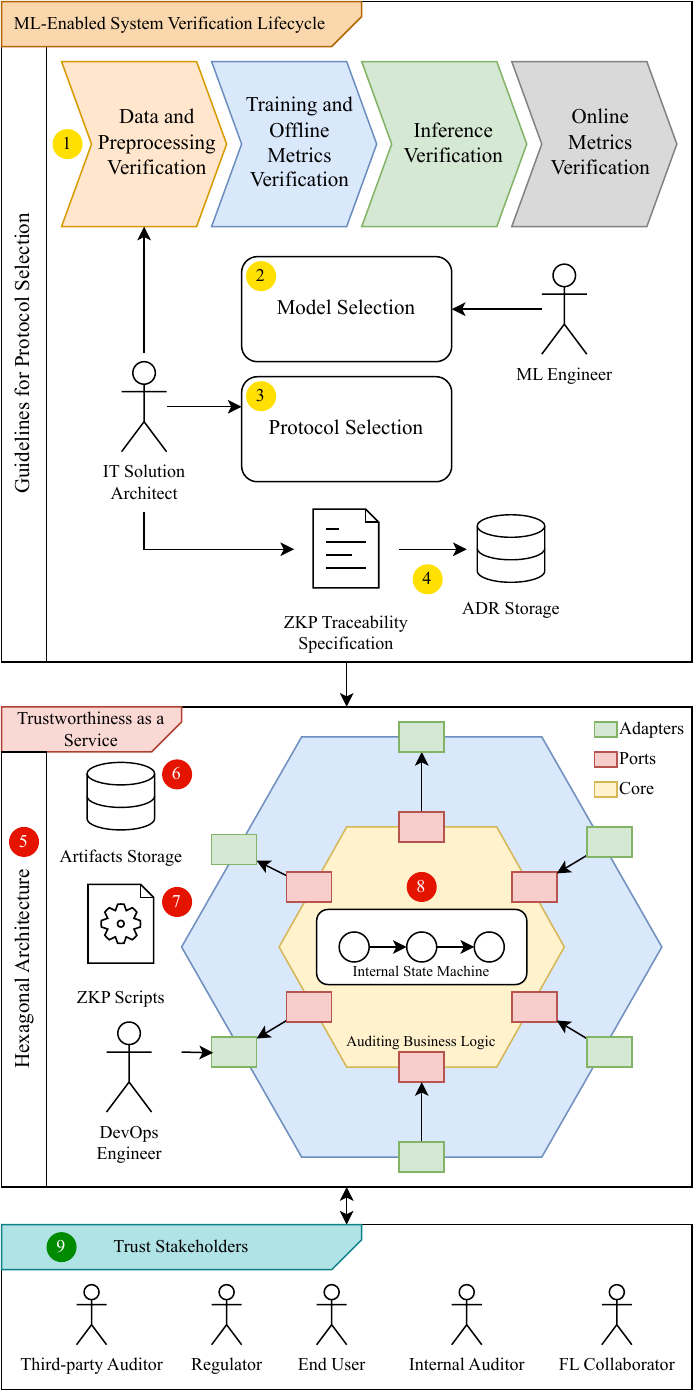}
    \caption{High-Level Overview of the ZKMLOps framework.}
    \label{fig:framework-overview}
\end{figure}

This section presents the ZKMLOps auditing framework and describes its building blocks, depicted in Figure~\ref{fig:framework-overview}.
The design lets \coloredcircle{9}{drawio_green}{white}\emph{Trust Stakeholders} verify specific properties across the MLOps verification lifecycle, serving as a mechanism for legal compliance assessment.
The framework sits at the intersection of two layers: a methodic part, with guidelines for operationalizing the first phase (\coloredcircle{1}{drawio_yellow}{black}, \coloredcircle{2}{drawio_yellow}{black}, \coloredcircle{3}{drawio_yellow}{black}, \coloredcircle{4}{drawio_yellow}{black}); and the practical part, comprising the \coloredcircle{5}{drawio_red}{white}implementation of a Hexagonal Architecture\cite{martin_clean_2018}, \coloredcircle{6}{drawio_red}{white}artifact storage, \coloredcircle{7}{drawio_red}{white}scripts executing the different ZKP protocols, and an \coloredcircle{8}{drawio_red}{white}internal state machine that tracks the auditing workflow and coordinates each script with its required artifacts.

\subsection{ML-Enabled System Verification Lifecycle: Guidelines for Protocol Selection}
\label{sec:guidelines}

This section describes the methodical part of the framework. At this stage, the IT Solution Architect (SA) collaborates with the ML Engineer (ML)~\cite{kreuzberger_machine_2023}. The output is a document describing (i) the purpose of the audit, (ii) the MLOps stage involved, (iii) the artifact to be audited (e.g., a model, a dataset, or a metric), and (iv) the chosen protocol. The last item depends on the previous ones, since each stage of the MLOps verification lifecycle has different implicit requirements, as do the implementation properties of the ML models.
Producing this document gives clear ownership and accountability over a specific audit practice.

\begin{figure}[!t]
\begin{tikzpicture}
\node[anchor=south west,inner sep=0] at (0,0) {\includegraphics[width=\linewidth]{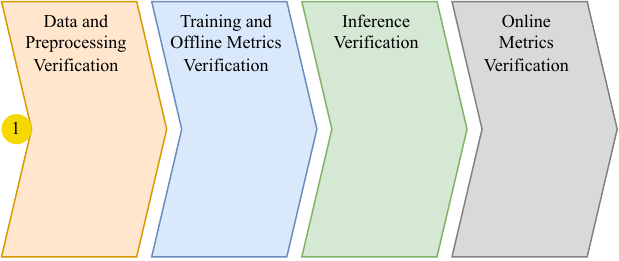}};
\draw (1.3,2.2) node{{
\cite{waiwitlikhit_trustless_2024}\cite{wang_efficient_2025}}};
\draw (3.4, 2.2) node{{ \cite{sun_zkdl_2025}\cite{zhao_veriml_2021}}};
\draw (3.4, 1.9) node{{
\cite{abbaszadeh_zero-knowledge_2024}\cite{waiwitlikhit_trustless_2024}}};
\draw (5.5, 2.2) node{{ \cite{zhang_zero_2020}\cite{liu_zkcnn_2021}}};
\draw (5.5, 1.9) node{{ \cite{ghaffaripour_mutually_2021}\cite{ju_efficient_2021}}};
\draw (5.5, 1.6) node{{ \cite{feng_zen_2021}\cite{zhao_veriml_2021}}};
\draw (5.5, 1.3) node{{
\cite{chen_zkml_2024}\cite{feng_zeno_2024}}};
\draw (5.5, 1) node{{ \cite{sun_zkllm_2024}\cite{wu_confidential_2024}}};
\draw (5.5, 0.7) node{{ \cite{lee_vcnn_2024}\cite{wang_efficient_2025}}};
\draw (7.7, 2.2) node{{ \cite{zhang_zero_2020}\cite{liu_zkcnn_2021}}};
\draw (7.7, 1.9) node{{ \cite{toreini_fairness_2024}\cite{wang_efficient_2025}}};
\end{tikzpicture}
\caption{ZKP-Enhanced ML applications in the MLOps verification lifecycle~\cite{scaramuzza_engineering_2025}.}
\label{fig:zkml_classification}
\end{figure}

In the first stage, the IT Solution Architect chooses a phase from the \coloredcircle{1}{drawio_yellow}{black}\emph{MLOps verification lifecycle} based on the purpose of the audit.
Scaramuzza et al.~\cite{scaramuzza_engineering_2025} propose a classification of ZKP-enhanced ML approaches from the white literature; each tackles a specific task of the MLOps verification lifecycle. The classification is depicted in Figure~\ref{fig:zkml_classification}. The four phases are derived from the \ac{TDSP} Model~\cite{amershi_software_2019}, with phases grouped for flexibility into: (i) \emph{Data and Preprocessing Verification}, (ii) \emph{Training and Offline Metrics Verification}, (iii) \emph{Inference Verification}, and (iv) Online Metrics Verification.
The Data and Preprocessing Verification phase includes the verification of properties related to dataset design choices and preprocessing operations, like model requirements, data collection, data cleaning and labeling, and feature engineering. The Training and Offline Metrics Verification includes the verification of the training process and the evaluation of model performance using metrics such as accuracy, F1-score, or group and individual fairness, which are computed right after the training and before the deployment, thus the name ``offline metrics''. This phase includes anything that fits into the Model Training and Model Evaluation phases. The Inference Verification phase focuses on ensuring the correctness of the inference computation process, and can happen only after the Model Deployment. Finally, the Online Metrics Verification phase involves the real-time verification of dynamic properties and metrics, such as model drift and live accuracy assessments, i.e., everything that fits into the Model Monitoring stage~\cite{scaramuzza_engineering_2025,amershi_software_2019}.

In the \coloredcircle{2}{drawio_yellow}{black}\emph{Model Selection} stage, the ML Engineer supports the IT Solution Architect by providing details about the deployed model and choosing the right verification technique based on its requirements. Scaramuzza et al.~\cite{scaramuzza_engineering_2025} identify nine categories of models that have been verified with ZKP in the literature. The results are shown in Table~\ref{tab:model_categories}.

\begin{table}[!t]
\centering
\scriptsize
\caption{ML Models Studied in the ZK-Enhanced ML Literature~\cite{scaramuzza_engineering_2025}.}
\label{tab:model_categories}
\begin{tabular}{@{}ll@{}}
\toprule
\coloredcircle{2}{drawio_yellow}{black}\textbf{ML Model Category} & \textbf{References} \\ \midrule
Decision Trees &
\cite{zhang_zero_2020}, \cite{zhao_veriml_2021} \\
Support Vector Machines &
\cite{ghaffaripour_mutually_2021}, \cite{zhao_veriml_2021}, \cite{wang_efficient_2025} \\
Linear Models (Linear/Logistic Regression) &
\cite{zhao_veriml_2021} \\
Clustering (K-Means) &
\cite{zhao_veriml_2021} \\
General Neural Networks &
\cite{feng_zen_2021}, \cite{feng_zeno_2024}, \cite{chen_zkml_2024},
\cite{zhao_veriml_2021}, \cite{wu_confidential_2024}, \cite{sun_zkdl_2025} \\
\acp{CNN} &
\cite{liu_zkcnn_2021}, \cite{ju_efficient_2021}, \cite{lee_vcnn_2024} \\
\acp{LLM} &
\cite{sun_zkllm_2024}\cite{chen_zkml_2024} \\
Vision Models &
\cite{chen_zkml_2024}, \cite{lee_vcnn_2024},
\cite{abbaszadeh_zero-knowledge_2024}, \cite{wu_confidential_2024},
\cite{waiwitlikhit_trustless_2024} \\
Recommender Systems (DLRM, Twitter) &
\cite{chen_zkml_2024}, \cite{waiwitlikhit_trustless_2024} \\
\bottomrule
\end{tabular}
\end{table}

These two phases can be summarised in a three-layered decision tree~\cite{magee1964decision}. The root node represents the audit purpose, the intermediate nodes the stages of the MLOps verification lifecycle, and the leaf nodes the models adopted. Each leaf lists a set of candidate protocols. At this point, the IT Solution Architect selects the protocol that best suits the application's architecture and organizational infrastructure. The literature may lack a specific implementation for a given audit purpose; in that case, the Architect considers alternative approaches. \coloredcircle{3}{drawio_yellow}{black}\emph{Protocol Selection} should maximize the number of properties a ZKP protocol holds for an ML application. These properties are (i) non-interactivity, (ii) transparent setup, (iii) standard representations, (iv) succinctness, and (v) post-quantum security~\cite{scaramuzza_engineering_2025}. Non-interactive protocols compress proof generation into a single exchange, where the prover submits a self-contained proof that any verifier
can check independently~\cite{canetti_fiat-shamir_2019}. 
The setup phase generates cryptographic parameters before proving begins. In a transparent setup, these parameters are derived solely from publicly available verifiable sources of randomness~\cite{sheybani_zero-knowledge_2025}, which suits applications requiring strong auditability and long-term trust guarantees. \ac{ZKP} protocols require computations to be expressed as formal representations compatible with their proof system~\cite{ernstberger_you_2024}. Standard, flexible representations broaden toolchain compatibility, developer accessibility, and the integration of ML models into various proof systems. Succinctness bounds both proof size and verifier time to poly-logarithmic in circuit size. Following Goldwasser, Kalai, and Rothblum~\cite{goldwasser_succinct_2008}, a non-interactive proof system is \emph{succinct} if both the proof size and the verifier time are bounded by $\textsc{poly}(\lambda + |x| + \log|C|)$, where $\lambda$ is the security parameter, $|x|$ is the size of the public input, and $|C|$ is the size of the computation. The verifier therefore runs in time \emph{poly-logarithmic} in $|C|$, which is what enables the millisecond-scale verification times reported in Section~\ref{sec:evaluation}. Whether a zero-knowledge protocol is considered post-quantum secure depends entirely on the primitives it employs. In general, protocols built solely on \acp{CRHF}~\cite{berman2018multi} are believed to be more resilient in a quantum context, since no quantum algorithm is currently known to break \acp{CRHF} faster than brute force. Conversely, protocols built on pairing-based polynomial commitment schemes (KZG~\cite{kate_constant-size_2010} as used in Halo2/ezkl, or the Groth16 setup over BN254 used by snarkjs) are \emph{not} post-quantum secure, because the discrete-logarithm problem they rely on is broken by Shor's algorithm~\cite{shor_algorithms_1994}. The Architect makes this trade-off explicit in the Traceability Specification: \ac{STARK}-style protocols (e.g., Cairo/Stwo) trade larger proof size and slower verification for hash-based, post-quantum-safe primitives, while pairing-based \acp{SNARK} deliver smaller proofs and faster verification at the cost of post-quantum security.

The output of this process is the \coloredcircle{4}{drawio_yellow}{black}\emph{ZKP Traceability Specification}, containing five fields: (i) The \textsc{Purpose of the Audit}, (ii) the \textsc{Decision Trace} (MLOps verification lifecycle phase and the deployed model), (iii) the \textsc{Selected Protocol}, (iv) a \textsc{Property Matrix} listing the five protocol properties (non-interactivity, transparent setup, standard representation, succinctness, post-quantum security) with a tick-or-cross for the chosen protocol, and (v) a one-paragraph \textsc{Trade-off Justification} explaining each property the Architect accepts $\times$ on. The Property Matrix and Trade-off Justification together operationalise the post-quantum criterion: a paper-trail that survives auditor review even decades later, when the protocol's quantum-security posture may need re-evaluation.
The ZKP Traceability Specification falls into the definition of an \ac{ADR}~\cite{LightweightArchitectureDecision}, and can be stored, following Richards et al.~\cite{richards_fundamentals_2025}, in an \emph{ADR Storage}, i.e., a wiki or a shared directory on a shared file server that can be accessed easily with a wiki or shared file server accessible to all actors involved.
\subsection{Trustworthiness-as-a-Service}
\label{sec:taas}

This section presents the building blocks of the application-oriented dimension of the framework. This dimension, named \emph{Trustworthiness as a Service}, is a service-oriented architecture that integrates arbitrary ZKP protocols into the MLOps auditing workflow.

The architectural pattern is the \emph{hexagonal architecture} \coloredcircle{5}{drawio_red}{white}, also known as \emph{onion} or clean architecture~\cite{martin_clean_2018}. It keeps the core business logic independent of external technology stacks, such as the various ZKP implementations and repositories. The operational logic uses a \textsc{Saga Pattern}~\cite{richardson_microservices_2018}: each audit workflow is decomposed into smaller procedural steps within an \emph{internal state machine}, and an orchestrator manages the program execution with a \textsc{State Pattern}~\cite{martin_clean_2018}.

\begin{figure}[!t]
    \centering
    \includegraphics[width=.9\linewidth]{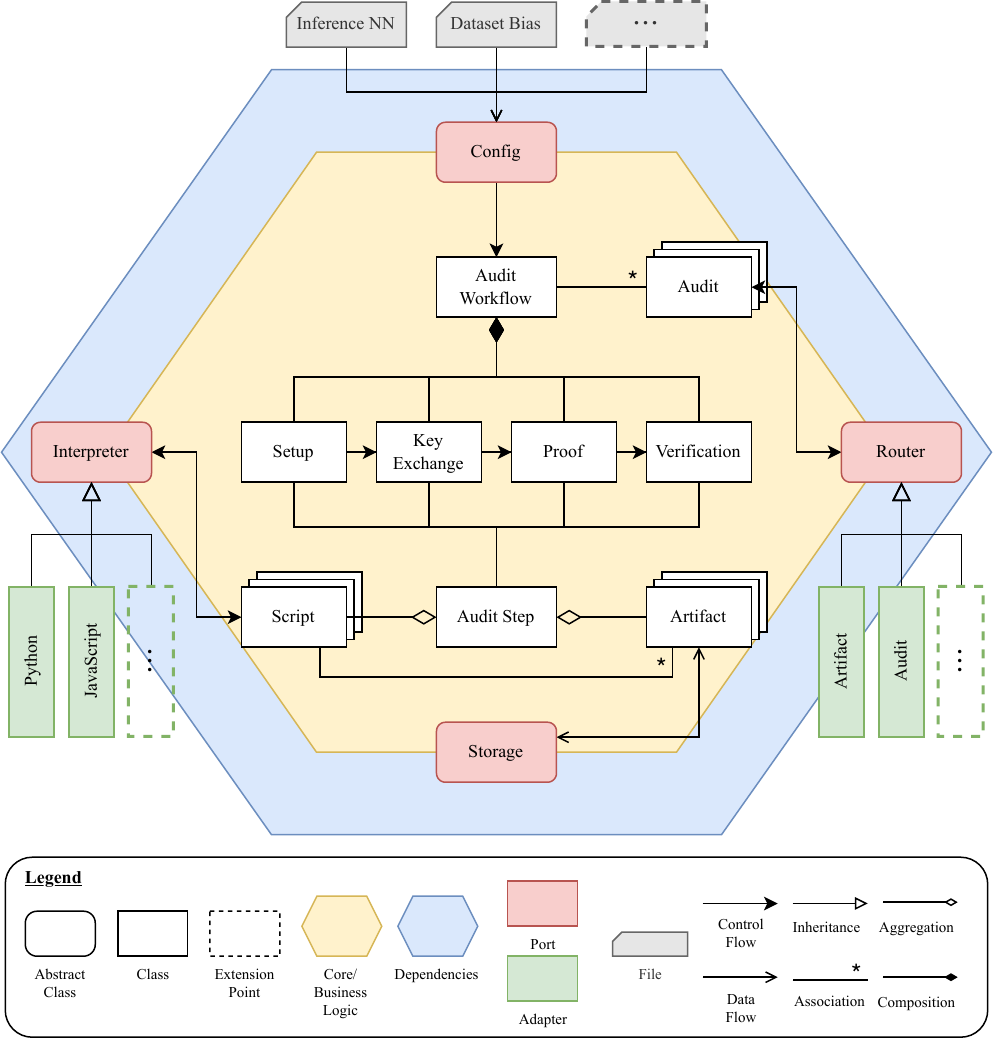}
    \caption{Component architecture of the ZKMLOps framework showing the implementation for two possible auditing workflows: neural network inference verification and dataset bias verification. The dotted boxes represent possible architecture extension points.}
    \label{fig:hexagon}
\end{figure}

In the hexagonal architecture, the system is divided into three main components: \emph{core}, \emph{ports}, and \emph{adapters}~\cite{martin_clean_2018}. The abstraction level increases toward the core: the inner layer contains higher-level policies and broad business rules, while the outermost layers handle concrete implementation details. 
Figure~\ref{fig:hexagon} shows the hexagonal architecture in detail. Adapters are implementations of ports, managed via a \textsc{Dependency Injection Pattern}~\cite{seemann_dependency_2012}, where adapters are ``injected'' at runtime if needed in the specific audit workflow.
The latter is defined through a configuration file and ingested through the \texttt{Config} port. In the Figure~\ref{fig:hexagon}, two scenarios are shown: neural network inference correctness verification and bias verification over a proprietary dataset. These configuration files are derived from the ZKP Traceability Specifications and contain the scripts needed to perform every step of the verification and the related input and output artifacts.

The inner hexagon, the \coloredhexagon{$\text{ }$}{drawio_styles_yellow_fill}{drawio_styles_yellow_border} \emph{core}, implements the business logic governing the auditing workflow. It orchestrates the program through four ZKP steps of an internal state machine \coloredcircle{8}{drawio_red}{white}: \emph{setup}, \emph{key exchange}, \emph{proof}, and \emph{verification}. The orchestrator combines an \textsc{Orchestrated Saga Pattern}~\cite{richardson_microservices_2018} with a \textsc{State Pattern}~\cite{gamma1995design}: each step checks a set of pre- and post-conditions to verify that input and output artifacts were correctly uploaded or generated in the artifact storage~\coloredcircle{6}{drawio_red}{white} and that the steps execute in order. 
Every running audit workflow is an \texttt{Audit} instance, i.e., a Saga.
Implementations and use cases vary, but the four steps are the common components in every non-interactive ZKP protocol. We abstracted these steps within the business logic and made the key-exchange step explicit: although it is conventionally folded into setup, separating it fits better in a service-oriented context, given the asynchronous nature of the auditing workflow. Each step is defined by a set of scripts that ingest and produce artifacts.

The outer hexagon, the \coloredhexagon{$\text{ }$}{drawio_styles_blue_fill}{drawio_styles_blue_border} \emph{dependencies}, implements an Anti-Corruption Layer against external dependencies~\cite{martin_clean_2018} using the \textsc{Ports and Adapters Pattern}.
The system defines one inbound port and three outbound ports. The inbound port, \texttt{Router}, lets the business logic be invoked by a set of REST APIs. This interface is implemented as dedicated adapters for each business logic (artifact management and audit management). The outbound ports are \texttt{Interpreter}, \texttt{Config}, and \texttt{Storage}. The first is the key component of the framework: it supports the execution of diverse protocols implemented as scripts~\coloredcircle{7}{drawio_red}{white} across varying technology stacks, via the \texttt{ScriptExecutor} service~\cite{senthilvel_enterprise_2017}, managed by the orchestrator with the \textsc{Visitor Pattern}~\cite{martin_clean_2018} and \textsc{Strategy Pattern}~\cite{martin_clean_2018}. The other two ports interface with the storage of configuration files and artifacts through a \textsc{Repository Pattern}~\cite{senthilvel_enterprise_2017}.

\subsection{Trust Stakeholders}
\label{sec:trust}

The \emph{Trust Stakeholders} \coloredcircle{9}{drawio_green}{white}are: \emph{third-party auditors}, \emph{regulators}, \emph{end users}, \emph{internal auditors}, or \emph{federated learning collaborators}. They interact with the system only through the interfaces exposed by the adapters. A GUI implemented as a web app would ease the auditing process.

\subsection{Trust and Threat Model}
\label{sec:trust-model}

\subsubsection{Roles and Trust Assumptions}
\label{sec:trust-roles}

It is important to note that ZKP-based auditing doesn't completely eliminate the need for trust, but it relocates it. Instead of requiring the verifier to trust the prover's honesty, the protocol narrows that trust to two foundations: the soundness of the cryptographic construction and the integrity of the artifact-binding mechanism.

The prover, i.e., the party whose pipeline artifact is under audit, is treated as untrusted with respect to the audited statement. The verifier, an Auditor or Regulator, is assumed honest-but-curious, the standard assumption for audit settings where the verifier has no incentive to forge proofs. For backends that require a trusted setup ceremony, such as KZG or Groth16, the setup authority is trusted only when the ceremony is publicly observable; in that case, the framework records the ceremony reference in the Traceability Specification. The host executing the scripts is trusted not to tamper with its own binaries; defending against a compromised host is outside the framework's scope. Artifact storage is treated as untrusted, with integrity enforced by the artifact commitment $H(M)$ described in Section~\ref{sec:trust-binding}. The configured ZKP backend is treated as trusted code; supply-chain attestation is delegated to standard \ac{SBOM} and signing practice.

\subsubsection{Artifact and Statement Binding}
\label{sec:trust-binding}

\paragraph*{\textit{Artifact Binding}}
The ZKP statement is bound to a cryptographic commitment to the artifacts that define the audited computation. The \texttt{setup} step computes $H(M)$ over a deterministically serialised bundle comprising the backend-specific compiled circuit representation, any calibration or configuration data the circuit depends on, and the parameter or lookup structures the prover must supply at proof time. For an inference workflow, this bundle concretely includes the compiled circuit (e.g., the \texttt{.ezkl} or \texttt{.r1cs} file), the post-calibration weight tensors, and any lookup tables for non-linear operations; for other workflow types, the bundle is defined by the \texttt{setup} script for that workflow. $H(M)$ is a public input to every subsequent proof. Any change to the bundle, whether re-compilation, recalibration, or a parameter update, breaks the binding and invalidates all prior proofs.

\paragraph*{\textit{Numerical Fidelity}}
The framework attests to the property of the \emph{quantised} circuit rather than the original floating-point model. Modern ZKML toolchains introduce up to $10^{-3}$ relative quantisation error~\cite{ezklGPU2023}; for compliance-grade audits the Architect must verify this is below the relevant regulatory threshold, increasing the scale factor or proving a separate error-bound circuit if needed.

\paragraph*{\textit{Input Binding}}
The Merkle root over the approved input set is a public input to the proof circuit; the prover submits an inclusion proof per audited execution~\cite{boneh2023graduate}. Range and policy proofs compose by adding linear constraints to the circuit. In the financial-risk use case, the approved-input set is the regulator-validated golden dataset; the resulting Merkle overhead is negligible relative to the proof circuit.

\subsubsection{Adversary Capabilities and Out-of-Scope Threats}
\label{sec:trust-threats}

\paragraph*{\textit{In-Scope Threats}} These threats include artifact bundle substitution (caught by $H(M)$ binding), input substitution outside the Merkle commitment, stale proof replay (defeated by binding the full public-input vector), and false-statement proofs (defeated by ZKP soundness).

\paragraph*{\textit{Out-of-Scope Threats}} These threats include: host compromise prior to setup, ZKP backend supply-chain attacks, side-channel leakage of witness values (zero-knowledge w.r.t.\ the protocol transcript, not physical observations), and denial of service against the orchestrator.

\subsubsection{Deployment Integrity Gap}
\label{sec:trust-deployment}

The framework's proofs attest to a property of the artifact bundle bound by $H(M)$, not to a property of the running production service. This is the \emph{deployment integrity} gap: a malicious deployer could keep committing to the audited artifact bundle while running a different one in production. Closing the gap requires runtime attestation, for example TEE-based remote attestation of the running service~\cite{costan2016intel}, signed transparency logs of artifact loads~\cite{rfc6962}, or a hash-chain of every served (input, output) pair tied back to $H(M)$. ZKMLOps and runtime attestation are complementary: ZKMLOps proves the \emph{property} of a committed artifact bundle; runtime attestation proves the \emph{identity} of the bundle in production. The same gap is discussed under Internal Validity in Section~\ref{sec:threats}.

\section{Example Use Case: Regulatory Compliance for a Proprietary Financial Risk Model}
\label{sec:use-case}

A Financial Institution must prove to an Auditing Company that the credit risk scores it produces via a machine learning model deployed as \ac{MLaaS} come from the declared, approved model. The model is proprietary and trained on sensitive data, so its internal parameters must remain confidential. Regulatory compliance still requires a verifiable process showing that the correct model was applied to approved input data without exposing the model itself.

After receiving the \textsc{Audit Request Letter} from the Auditing Company \coloredcircle{9}{drawio_green}{white}, the IT Solution Architect and the Machine Learning Engineer choose the protocol for the audit workflow \coloredcircle{1}{drawio_yellow}{black}\coloredcircle{2}{drawio_yellow}{black}\coloredcircle{3}{drawio_yellow}{black}. A simplified version of the resulting ZKP Traceability Specification is shown below:

\begin{framed}
  \begin{itemize}
      \item \textbf{\textsc{Purpose of the Audit:}}
      Verify that the financial institution's deployed neural network model correctly computes credit risk scores on approved input data during the inference phase, without exposing proprietary model parameters.

      \item \textbf{\textsc{Decision Trace:}}
      Focus on the inference phase of the MLOps lifecycle. The approach ensures the deployed neural network's outputs can be proven correct while maintaining confidentiality.

      \item \textbf{\textsc{Selected Protocol:}}
      \texttt{ezkl} (Halo2 over the BN254 curve with KZG polynomial commitments). The protocol is non-interactive, has a universal but non-transparent setup (the Halo2 KZG ceremony), standard representation, and succinctness. It is \emph{not} post-quantum secure: KZG relies on the hardness of pairing-based discrete logarithms on BN254, which Shor's algorithm breaks in polynomial time on a sufficiently large quantum computer~\cite{shor_algorithms_1994}. Post-quantum security is therefore reported as $\times$ in the Traceability Specification; the Architect accepts this trade-off in exchange for the proof-size and verification-time advantages measured in Section~\ref{sec:evaluation}.

      \item \textbf{\textsc{Property Matrix:}} non-interactive $\checkmark$, transparent setup $\times$ (Halo2 KZG ceremony), standard representation $\checkmark$ (\ac{ONNX} $\rightarrow$ Halo2 circuit), succinctness $\checkmark$, post-quantum $\times$.

      \item \textbf{\textsc{Trade-off Justification:}} The audit is a one-shot regulatory inquiry, not a long-term archival proof; pairing-based primitives are accepted because the proof's relevance horizon (5--10 years per Basel-style audit cycles) is shorter than current quantum-computer capability projections. Should this assessment change, the framework allows re-running the audit under a STARK backend without changes to the orchestration layer.
  \end{itemize}
\end{framed}

\begin{figure}[!t]
    \centering
    \includegraphics[width=\linewidth]{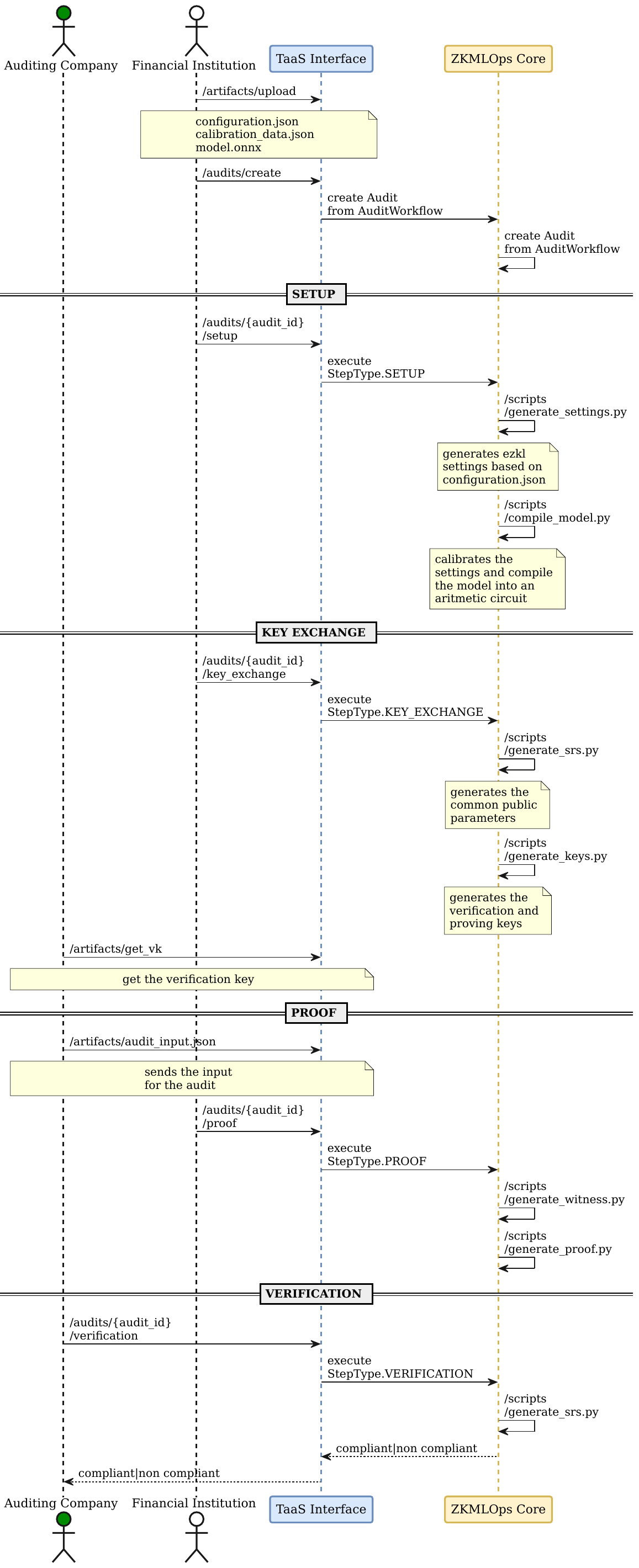}
    \caption{Sequence Diagram of the Neural Network Inference Audit Workflow.}
    \label{fig:seq_diagram}
\end{figure}

The workflow follows the sequence diagram shown in Figure~\ref{fig:seq_diagram}.

If the output is correctly generated by the declared model, the audit workflow concludes with a response \textcolor{drawio_green}{\texttt{compliant}}; otherwise, it shall return \textcolor{drawio_red}{\texttt{non compliant}}.

The Financial Institution shares only the verification key generated during the \texttt{Key Exchange} step and the proof produced in the \texttt{Proof} step; the Merkle root of the approved-input dataset and the model commitment $H(M)$ are also included as public inputs to the proof, as described in Section~\ref{sec:trust-binding}. Nothing else is disclosed, preserving confidentiality and protecting intellectual property.

Once the verifier accepts the proof, the auditor holds cryptographic evidence that the inference was produced by the model committed in the Traceability Specification. The auditor combines this evidence with the rest of the regulatory checklist (provenance of the approved input set, alignment between the audited statement and the regulator's required property, deployment-integrity attestation per Section~\ref{sec:trust-deployment}) before issuing a compliance certificate. The Financial Institution, in turn, doesn't disclose any confidential information or intellectual property.

\section{Evaluation}
\label{sec:evaluation}

\subsection{Setup}
\label{sec:eval-setup}

All measurements were collected on an Apple M4~Pro workstation with 48~GB of unified memory running macOS~15, using arm64-native toolchains. Each repetition was preceded by an unrecorded warmup to prime the file-system cache; repetitions executed sequentially with a 5~s gap; any repetition whose CPU frequency dropped below 95\% of the nominal P-core frequency was discarded as thermally throttled. For ezkl-based experiments we report mean and standard deviation across $N{=}10$ repetitions. For the snarkjs experiment we report $N{=}3$, which bounds the variance of the framework-overhead measurement below 2\%. The MobileNet-v2 sweep runs at $N{=}1$ per variant, since its purpose is to characterise stable end-to-end behaviour across circuit sizes rather than per-variant variance.

\subsection{Framework Viability (RQ1)}
\label{sec:eval-rq1}

To answer \textbf{RQ1} we evaluate the framework along two complementary dimensions: (i)~cross-backend abstraction on small models, where the orchestration cost is most visible relative to short proving times, and (ii)~stability as the cryptographic workload grows, evaluated through a sweep of progressively larger MobileNet-v2 variants. Table~\ref{tab:models} summarises the models used.

\begin{table*}[t]
\centering
\caption{Benchmark model lineup. The lineup pairs two small tabular models, evaluated under both ezkl (Halo2/KZG) and snarkjs (Groth16), with a five-point MobileNet-v2 sweep used as the framework's stability stress test (ezkl only).}
\label{tab:models}
\begin{tabular}{llrrrl}
\toprule
\textbf{Model} & \textbf{Dataset} & \textbf{Params} & \textbf{MACs} & \textbf{Input shape} & \textbf{Protocols} \\
\midrule
  Logistic Regression & UCI Adult Income & 110 & 108 & 1×108 & ezkl, snark \\
  MLP-Tabular & UCI German Credit & 7.8K & 7.7K & 1×24 & ezkl \\
  \midrule
  MobileNet-v2 0.35@96 & CIFAR-10 (96×96) & 415.0K & 60.00M & 1×3×96×96 & ezkl \\
  MobileNet-v2 0.50@128 & CIFAR-10 (128×128) & 720.0K & 165.00M & 1×3×128×128 & ezkl \\
  MobileNet-v2 0.75@160 & CIFAR-10 (160×160) & 1.48M & 470.00M & 1×3×160×160 & ezkl \\
  MobileNet-v2 1.00@192 & CIFAR-10 (192×192) & 2.25M & 1040.00M & 1×3×192×192 & ezkl \\
  MobileNet-v2 1.00@224 & CIFAR-10 (224×224) & 3.50M & 1820.00M & 1×3×224×224 & ezkl \\
\bottomrule
\end{tabular}
\end{table*}

\paragraph*{\textit{Models and Rationale}} We deliberately use two small tabular models as our primary subjects. When proving completes in under a second, every millisecond spent outside the prover is exposed; this is the regime in which any hidden orchestration cost is hardest to hide. Logistic Regression on UCI Adult Income is the only model exercised through both ezkl (Halo2/KZG, Python-native) and snarkjs (Groth16, Node.js CLI), and acts as our cross-backend anchor. The \ac{MLP}-Tabular model on UCI German Credit corresponds to the financial-risk use case of Section~\ref{sec:use-case}. The five-point MobileNet-v2 sweep on ezkl, ranging from $0.4$M to $3.5$M parameters, drives proving time from sub-second to multi-second magnitudes, letting us observe whether orchestration remains responsive at scale.

\paragraph*{\textit{Measuring Framework Overhead}} For each audit, the orchestrator records the total wall-clock time $T_{\textit{total}}$ observed end-to-end and, for each sub-step $i$, the time $T^{\textit{self}}_{i}$ spent inside the cryptographic script itself. The framework overhead is the share of total time spent \emph{outside} cryptographic primitives: HTTP round-trips between the harness and the orchestrator, subprocess spawning, audit bookkeeping, and inter-step pre/post-condition checks:
\begin{equation}
\textit{framework\_overhead\_pct} = \frac{T_{\textit{total}} - \sum_{i} T^{\textit{self}}_{i}}{T_{\textit{total}}}\cdot 100
\label{eq:framework-overhead}
\end{equation}

\begin{table*}[t]
\centering
\caption{Protocol benchmark for the experiments driven end-to-end through the orchestrator. \textit{Prove}, \textit{Verify}, and \textit{Overhead} are reported as mean~$\pm$~std across the repetition set ($N{=}10$ for ezkl rows, $N{=}3$ for snarkjs). \textit{Overhead} is the framework cost defined in Equation~\ref{eq:framework-overhead}.}
\label{tab:protocol_bench}
\begin{tabular}{llrrrr}
\toprule
\textbf{Model} & \textbf{Protocol} & \textbf{Prove (ms)} & \textbf{Verify (ms)} & \textbf{Proof (KB)} & \textbf{Overhead (\%)} \\
\midrule
  Logistic Regression & EZKL & $607 \pm 31$ & $155 \pm 45$ & 33.2 & $37.8 \pm 2.8$ \\
   & SNARK & $409 \pm 34$ & $326 \pm 10$ & 20.7 & $32.6 \pm 1.3$ \\
  \midrule
  MLP-Tabular & EZKL & $739 \pm 19$ & $151 \pm 27$ & 20.8 & $29.1 \pm 1.0$ \\
\bottomrule
\end{tabular}
\end{table*}

\paragraph*{\textit{Cross-Backend Abstraction}} The rightmost column of Table~\ref{tab:protocol_bench} carries the central viability claim. On Logistic Regression, the orchestrator imposes $37.8 \pm 2.8\%$ overhead on the ezkl pipeline and $32.6 \pm 1.3\%$ on the snarkjs pipeline. The two backends differ substantially in architecture (a Python-native library on one side, a Node.js command-line tool on the other), yet their overheads land within five percentage points of each other, with sub-3\% standard deviation. We read this as evidence that the abstraction is real and not specific to a single backend. On MLP-Tabular under ezkl, overhead drops further to $29.1 \pm 1.0\%$, consistent with the orchestrator imposing a roughly constant absolute cost per audit that is amortised over a slightly larger cryptographic denominator. The remaining columns (prove time, verify time, and proof size) are not intended as a backend ranking; they reproduce the well-known protocol-level trade-offs (Halo2's faster verification, Groth16's smaller proofs) and confirm that the orchestrator exposes the underlying primitives without masking them.

\begin{figure}[!t]
  \centering
  \includegraphics[width=\linewidth]{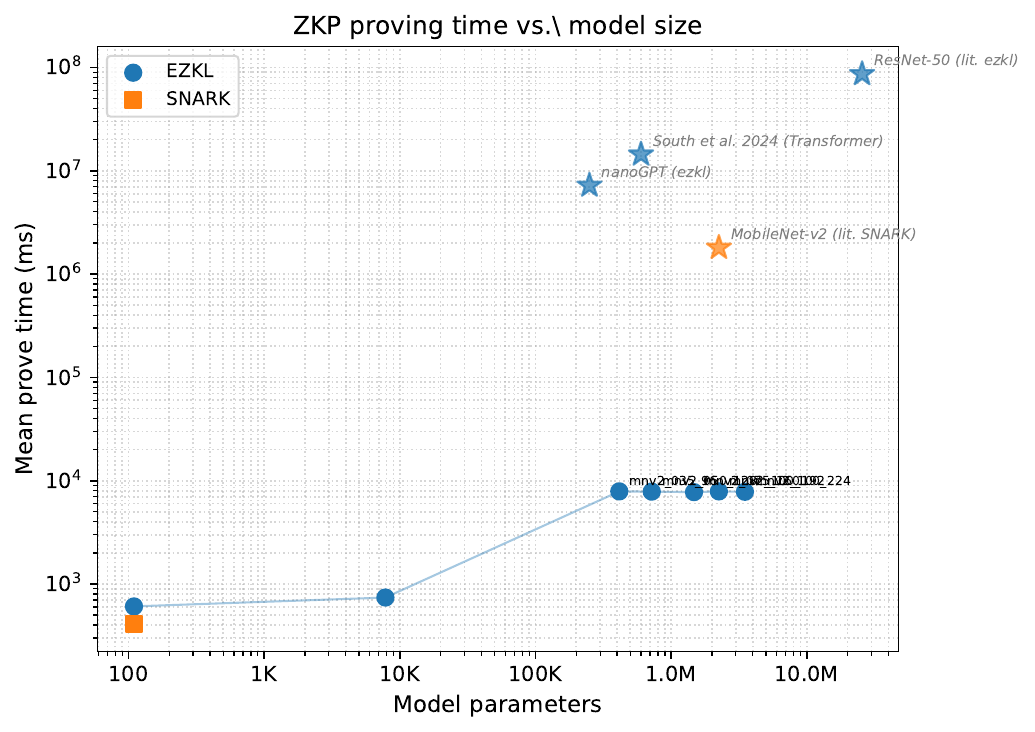}
  \caption{Proving time vs.\ model size for the experiments driven end-to-end through the orchestrator (small tabular models for ezkl and snarkjs; the MobileNet-v2 sweep for ezkl). Literature anchor points are overlaid for context.}
  \label{fig:scaling-provetime}
\end{figure}

\paragraph*{\textit{Behaviour Under Increasing Circuit Size}} Table~\ref{tab:scaling} and Figure~\ref{fig:scaling-provetime} report the MobileNet-v2 sweep, in which the cryptographic step dominates wall-clock time. All five variants completed end-to-end through the orchestrator within the 44~GB memory ceiling imposed on the 48~GB host. Per-repetition prove time is nearly flat across the sweep ($7745$--$7877$~ms), as the same configuration absorbs the entire span from $0.4$M to $3.5$M parameters with only ${\approx}2\%$ variation in prove time. The cause is ezkl's calibration step, which assigns each model a \emph{logrows} value $K$ and pads the arithmetic circuit to exactly $2^{K}$ rows, regardless of actual utilisation. The dominant proof operation, a multi-scalar multiplication (MSM) over the KZG structured-reference string, scales as $O(2^{K})$ and is therefore insensitive to how many of those rows a given model occupies. All five sweep variants were assigned the same $K$ by the auto-calibration, so their MSM workloads are identical and their prove times are indistinguishable within measurement noise. The relevant observation for the framework is that the ${\sim}500$~ms of HTTP and subprocess cost measured on the small models does not grow with circuit size; as the cryptographic workload grows, the orchestration share becomes a vanishing fraction of total wall-clock.

\begin{table}[t]
\centering
\caption{Framework stability under increasing circuit size: MobileNet-v2 sweep driven through the orchestrator with the ezkl backend.}
\label{tab:scaling}
\begin{tabular}{lrlr}
\toprule
\textbf{Variant} & \textbf{Params} & \textbf{Outcome} & \textbf{Prove (ms)} \\
\midrule
  MobileNet-v2 0.35@96 & 415K & ok & 7877 \\
  MobileNet-v2 0.50@128 & 720K & ok & 7832 \\
  MobileNet-v2 0.75@160 & 1475K & ok & 7745 \\
  MobileNet-v2 1.00@192 & 2250K & ok & 7872 \\
  MobileNet-v2 1.00@224 & 3500K & ok & 7803 \\
\bottomrule
\end{tabular}
\end{table}

\begin{framed}
\noindent\textbf{Finding (RQ1).} The ZKMLOps orchestrator imposes a bounded, backend-invariant overhead (within five percentage points across architecturally distinct ZKP implementations) that becomes a vanishing fraction of total wall-clock time as model and circuit size grow.
\end{framed}

\subsection{Comparative Baselines (RQ2)}
\label{sec:eval-rq2}

The viability result of Section~\ref{sec:eval-rq1} does not, on its own, say when full zero-knowledge auditing is the appropriate tool. Full \acp{ZKP} offer confidentiality of weights and inputs, succinct verification, and replayable evidence, without trusting hardware vendors or assuming non-collusion among compute parties. They are also several orders of magnitude slower than plain inference. RQ2 places these full-ZK results next to the alternative verification techniques that recur in the literature, so the cost-versus-assurance trade-off becomes explicit.

We compare three families of techniques: plain inference (without any integrity guarantee), sampling combined with verifiable computation, and trusted execution environments (TEE) or secure multi-party computation (MPC). Two configurations were measured on the same hardware that produced Section~\ref{sec:eval-rq1}: a plain ONNX Runtime baseline that anchors the prover-overhead column at $1\times$, and a sampling harness that drives the same orchestrator at sampling rates $\alpha \in \{10^{-3}, 10^{-2}, 10^{-1}\}$ over a stream of 300 inputs, with three repetitions per configuration. The remaining rows are reported from the literature with explicit hardware footnotes: our host does not have SGX-2, and there is not a clean arm64-macOS path for the cited MPC frameworks. Table~\ref{tab:baselines} summarises the comparison; the full-ZK rows reuse figures from Section~\ref{sec:eval-rq1}.

\begin{table*}[t]
\centering
\caption{Comparative baselines for RQ2. \textit{Measured} rows were collected on the same machine that produced Table~\ref{tab:protocol_bench}; \textit{cited} rows carry per-row hardware footnotes. Prover overhead is reported as a multiple of plain ONNX-Runtime CPU inference on the same host where \textit{measured}, and as the literature value when \textit{cited}. \textit{Comm.} is the communication size of the artifact exchanged between the prover and the verifier. \textit{Conf. wts} and \textit{Conf. ins} are the confidentiality of the weights and of the inputs, respectively. \textit{PQ} is post-quantum security. }
\label{tab:baselines}
\resizebox{\linewidth}{!}{%
\begin{tabular}{p{3.4cm}p{2.6cm}rrrcccl}
\toprule
\textbf{Technique} & \textbf{Trust assumption} & \textbf{Prover ovh.} & \textbf{Verify} & \textbf{Comm.} & \textbf{Conf. wts} & \textbf{Conf. ins} & \textbf{PQ} & \textbf{Source} \\
\midrule
  Plain inference, Logistic Regression & trust prover fully & $1\times$ (4.7~\textmu s) & n/a & 0 & $\times$ & $\times$ & n/a & measured \\
  Plain inference, MLP-Tabular & trust prover fully & $1\times$ (6.7~\textmu s) & n/a & 0 & $\times$ & $\times$ & n/a & measured \\
  Sampling+VC, Logistic Regression, $\alpha{=}10^{-3}$ & crypto + sampling & $317\times$ amort.\ (1.49~ms) & per-sample ZK + chain & per-sample proof & \checkmark (sampled) & \checkmark (sampled) & ezkl: KZG & measured \\
  Sampling+VC, Logistic Regression, $\alpha{=}10^{-2}$ & crypto + sampling & $\sim 10^{3}\times$ amort.\ (5.71~ms) & per-sample ZK + chain & per-sample proof & \checkmark (sampled) & \checkmark (sampled) & ezkl: KZG & measured \\
  Sampling+VC, Logistic Regression, $\alpha{=}10^{-1}$ & crypto + sampling & $\sim 10^{4}\times$ amort.\ (58.79~ms) & per-sample ZK + chain & per-sample proof & \checkmark (sampled) & \checkmark (sampled) & ezkl: KZG & measured \\
  Sampling+VC, MLP-Tabular, $\alpha{=}10^{-3}$ & crypto + sampling & $0.72\times$ amort.\ (4.8~\textmu s) & per-sample ZK + chain & per-sample proof & \checkmark (sampled) & \checkmark (sampled) & ezkl: KZG & measured \\
  Sampling+VC, MLP-Tabular, $\alpha{=}10^{-2}$ & crypto + sampling & $\sim 10^{3}\times$ amort.\ (7.89~ms) & per-sample ZK + chain & per-sample proof & \checkmark (sampled) & \checkmark (sampled) & ezkl: KZG & measured \\
  Sampling+VC, MLP-Tabular, $\alpha{=}10^{-1}$ & crypto + sampling & $\sim 10^{4}\times$ amort.\ (70.86~ms) & per-sample ZK + chain & per-sample proof & \checkmark (sampled) & \checkmark (sampled) & ezkl: KZG & measured \\
  Full ZK, ezkl (Halo2/KZG), Logistic Regression & crypto only & $\sim 10^{5}\times$ (594~ms) & 135~ms & 33.2~KB & \checkmark & \checkmark & $\times$ (KZG) & Table~\ref{tab:protocol_bench} \\
  Full ZK, snarkjs (Groth16), Logistic Regression & crypto only & $\sim 10^{4}\times$ (390~ms) & 327~ms & 20.7~KB & \checkmark & \checkmark & $\times$ (Groth16) & Table~\ref{tab:protocol_bench} \\
  Full ZK, ezkl (Halo2/KZG), MLP-Tabular & crypto only & $\sim 10^{5}\times$ (732~ms) & 138~ms & 20.8~KB & \checkmark & \checkmark & $\times$ (KZG) & Table~\ref{tab:protocol_bench} \\
  \midrule
  TEE / SGX (Slalom), VGG-16 & trust HW + microcode & 1.05--2$\times$ & ms (attestation) & KB & \checkmark & \checkmark & $\times$ & \cite{tramer_slalom_2019} \\
  TEE / SGX (Occlumency), AlexNet & trust HW + microcode & 1.22--2.13$\times$ & ms (attestation) & KB & \checkmark & \checkmark & $\times$ & \cite{lee_occlumency_2019} \\
  TEE / SGX-2, capability note & trust HW + microcode (Ice Lake server) & n/a & n/a & n/a & \checkmark & \checkmark & $\times$ & \cite{elhindi_sgx2_2022} \\
  MPC (semi-honest 2PC), ResNet-50 (ImageNet) & non-collusion 2PC & 30~s wall & = prover (interactive) & GBs & \checkmark & \checkmark & \textrm{---} & \cite{rathee_cryptflow2_2020} \\
  MPC (Piranha, GPU), VGG-16 & non-collusion 3PC & 1.6 (inf.) & = prover (interactive) & GBs & \checkmark & \checkmark & \textrm{---} & \cite{watson_piranha_2022} \\
  MPC (DELPHI), ResNet-32 & semi-honest 2PC & \textrm{---} & online 22x faster than Gazelle & MBs--GBs & \checkmark & \checkmark & \textrm{---} & \cite{mishra_delphi_2020} \\
  Auditing (Arc) & crypto-only commitments & up to $10^4\times$ faster vs.\ hash-based & ms & KB & \checkmark & \checkmark & \textrm{---} & \cite{lycklama_arc_2024} \\
  Sampling+SMPC (Priveri), Llama-2-7B (1 token) & SMPC + sentinels & 22.1~s wall & per-sample & MBs & \checkmark & \checkmark & \textrm{---} & \cite{pal_priveri_2024} \\
  Sampling (FaaS), per-user cryptograms & per-user cryptograms & \textrm{---} & per-user & KB & $\times$ & \checkmark & \textrm{---} & \cite{toreini_fairness_2024} \\
  Sampling+VC (IMMACULATE), design only & crypto + sampling at rate $\alpha$ & amortised by $1/\alpha$ & per-sample ZK + chain & per-sample proof & \checkmark & \checkmark & \textrm{---} & \cite{guo_immaculate_2024} \\
\bottomrule
\end{tabular}%
}
\end{table*}

\paragraph*{\textit{Plain Inference}} The plain-inference rows establish the lower bound: no cryptographic cost (microseconds per inference) but no integrity or confidentiality guarantees, since the auditor must trust the prover entirely. They are useful only as the anchor for the prover-overhead column.

\paragraph*{\textit{Sampling with Verifiable Computation}} This baseline targets streaming inference, where steady-state cost per request matters more than per-request integrity. A fraction $\alpha$ of inferences carries a full ZK proof; the remainder is bound to a hash-chain root that the auditor can spot-check. The amortised cost per inference is $\alpha \cdot \texttt{prove\_ms} + (1-\alpha) \cdot \texttt{plain\_ms}$. On Logistic Regression, dropping $\alpha$ from $10^{-1}$ to $10^{-3}$ reduces the amortised cost from ${\sim}59$~ms to ${\sim}1.5$~ms, four orders of magnitude below the full-ZK row in the same table; MLP-Tabular behaves the same way. The trust assumption weakens accordingly: only the sampled inferences carry confidentiality and integrity guarantees, while the rest of the stream is covered only by the chain root. This is a useful operating point for telemetry-style auditing, but not a replacement for full ZK in the audit-on-demand setting that motivates the framework.

\paragraph*{\textit{Trusted Execution Environments (TEE)}} The cited TEE rows (Slalom, Occlumency) place prover overhead in the $1.05$--$2.13\times$ band on VGG-16 and AlexNet, with kilobyte-scale communication and millisecond-scale verification. Two caveats narrow the applicability of this regime. First, \ac{SGX}-1 enforces a 90--256~MB \ac{EPC} ceiling, which Occlumency's AlexNet measurement already exceeds, so the cited overheads partly reflect \ac{EPC} paging rather than the underlying primitive; \ac{SGX}-2 largely removes this bottleneck on Ice~Lake server hardware, which is not present on our test host. Second, the trust assumption is strictly stronger than the crypto-only one: a microcode bug or a side-channel attack on the enclave breaks both confidentiality and integrity in ways that cryptographic protocols are designed to resist.

\paragraph*{\textit{Multi-Party Computation}} The cited MPC rows (CryptFlow2, Piranha, DELPHI) avoid hardware trust but require a non-collusion assumption among compute parties, communication in the megabyte to gigabyte range per inference, and an interactive protocol that ties prover and verifier together at runtime. This is the opposite of the offline, replayable evidence that the framework targets.

\paragraph*{\textit{When Full ZK is the Right Tool}} The audit-on-demand setting is the regime in which the full-ZK results of Table~\ref{tab:baselines} are the right answer. Each audit verifies a single inference; the auditor is offline at proof generation; verification must be succinct, non-interactive, and replayable months later from cold storage. The full-ZK rows carry confidentiality of weights and inputs without trusting hardware or non-collusion among compute parties, at sub-second prover cost on tabular models and verifier cost in the 135--327~ms band. The MobileNet-v2 sweep of Section~\ref{sec:eval-rq1} shows that the same orchestration profile carries through to circuits of millions of parameters, so the regime is not confined to small tabular models. The hash-based commitment baseline (Arc) is reported as a related point of comparison: it sits next to the full-ZK regime in the trust model but trades verifier succinctness for prover speedup, a dimension we do not measure here.

\begin{framed}
\noindent\textbf{Finding (RQ2).} Full zero-knowledge auditing is the appropriate tool for audit-on-demand settings requiring crypto-only trust and replayable evidence; sampling-with-VC and TEE-based approaches are preferable for high-throughput streaming and hardware-trust settings respectively.
\end{framed}

\section{Discussion}
\label{sec:discussion}

ZKMLOps resolves the tension between audit transparency and asset confidentiality by giving verifiers cryptographic evidence about audited computations without granting access to the underlying data or model.

\paragraph*{\textit{Answer to RQ1}} Tables~\ref{tab:protocol_bench} and~\ref{tab:scaling} confirm that the orchestrator imposes a tightly bounded, backend-invariant overhead (within five percentage points across ezkl and snarkjs) that becomes a vanishing fraction of wall-clock as circuit size grows. GPU-accelerated proving, available in ezkl~\cite{ezklGPU2023}, is a path toward larger-model support.

\paragraph*{\textit{Answer to RQ2}} Table~\ref{tab:baselines} shows that the appropriate technique depends on the trust model and workload shape: full ZK is the right tool for audit-on-demand settings (succinct, non-interactive, crypto-only trust), while sampling-with-VC and TEE suit streaming and hardware-trust settings respectively.

\paragraph*{\textit{Design Principles}} The main SE finding is that ZKP primitives can be integrated into MLOps via four collaborating patterns: \textsc{Hexagonal Architecture} (macro-level structure), \textsc{Orchestrated Saga} (business process), \textsc{State} (workflow steps), and \textsc{Strategy} (interchangeable script behaviors). This hierarchy decouples core auditing logic from external dependencies, so a new ZKP backend or audit type can be added without changing the orchestration. The broader claim that ZKMLOps enables governance throughout the MLOps lifecycle remains aspirational and is addressed in future work.

\section{Implications for Practice}
\label{sec:implications}

\paragraph*{\textit{ML Engineers and DevOps Teams}} ZKMLOps integrates incrementally alongside existing MLOps pipelines. The ZKP Traceability Specification, which records audit purpose, MLOps phase, selected protocol, and trade-off justification, acts as an Architecture Decision Record stored in a shared repository, making cryptographic design choices traceable and reviewable by the broader team. Teams can begin with the inference-verification workflow, the most mature in the current ZKML ecosystem, and extend to training or data-preprocessing audits as toolchain support grows. The hexagonal adapter interface isolates these extensions from the core orchestration logic, so adopting a new ZKP backend requires no changes to the surrounding pipeline.

\paragraph*{\textit{Compliance Officers and Auditors}} The evidence bundle produced by ZKMLOps (a proof, a verification key, a model commitment $H(M)$, and a Merkle root over approved inputs) provides a cryptographically verifiable binding between a claimed computational property and the audited artifact bundle. Auditors should note that this binding attests to the \emph{committed artifact}, not to the running production service; closing the deployment integrity gap requires complementary runtime attestation (e.g., TEE-based remote attestation or signed artifact logs), which the framework documents explicitly in the Traceability Specification (Section~\ref{sec:trust-deployment}). The resulting audit trail is designed to align with evidence requirements imposed by regulatory frameworks such as the EU AI Act.

\paragraph*{\textit{Tool and Framework Developers}} STARK-based backends offer transparent setup and post-quantum security, making them architecturally preferable for long-horizon archival audits where proof validity must survive beyond current quantum-computing projections. The main bottleneck for large models is proving time; GPU-accelerated proving, already available in ezkl~\cite{ezklGPU2023}, is the most tractable path toward full-ZK auditing of larger architectures. New ZKP backends integrate with ZKMLOps by implementing the hexagonal adapter port interface and defining setup, key-exchange, proof, and verification scripts together with a configuration file, without modifying the core orchestration layer.

\section{Threats to Validity and Limitations}
\label{sec:threats}

We address four threats to validity following Feldt et al.~\cite{feldt_validity_nodate}: \emph{external validity}, \emph{conclusion validity}, \emph{internal validity}, and \emph{construct validity}.

\textbf{External Validity.}
We show the framework through an illustrative use case in financial risk auditing. The ZKMLOps architecture may not transfer directly to other regulated fields, such as healthcare or autonomous systems, which have different compliance requirements and operational demands. The hexagonal architecture decouples the core auditing logic from domain-specific implementations, limiting this threat.

\textbf{Conclusions Validity.}
The evaluation focuses on inference verification, omitting other phases of the MLOps lifecycle. The main reason is that inference is the most mature sub-field of \ac{ZKML}, providing a reliable baseline for assessing the stability of the framework's orchestration. Moreover, the modularity inherent in the hexagonal architecture makes these findings generalizable, as a new audit workflow can be integrated with new adapters without altering the core logic of the orchestration layer. 

\textbf{Internal Validity.}
The observed performance metrics could reflect the maturity and optimization level of the underlying cryptographic libraries rather than the theoretical efficiency of the protocols themselves. We addressed this threat by running all benchmarks in a standardized experimental environment, holding hardware and operating system variables constant across tests. Averaging over multiple random initializations reduced transient system noise, giving a comparison of practical tool performance rather than a purely theoretical one. We further note the deployment-integrity gap discussed in Section~\ref{sec:trust-deployment}: our measurements bind a committed artifact, not a running service; closing this gap requires runtime attestation outside the framework's scope.

\textbf{Construct Validity.}
The theoretical construct of trustworthiness includes multiple attributes, whereas our evaluation is confined to verifying the computational correctness of model inference, a specific aspect of auditing. We mitigate this threat by scoping our contribution explicitly. The core value of ZKMLOps is the implementation of its orchestration layer, which manage the complex lifecycle of \acp{ZKP} transforming them into unified and repeatable workflows. While other dimensions of trustworthiness were not the primary focus of this work, the framework's architecture ensures these constructs can be integrated as extensions.

\section{Conclusions and Future Work}
\label{sec:conclusions}

ZKMLOps is a framework for auditing ML systems under the accountability requirements imposed by regulation. Zero-Knowledge Proofs provide the technical foundation: a prover convinces a verifier of a computational claim without revealing the underlying data or model, producing \emph{verifiable evidence} that auditors can fold into a compliance assessment.

The framework maps ZKP steps onto an orchestrated, adapter-based lifecycle, so audits are modular and repeatable across heterogeneous backends. We showed its practicality through a financial risk auditing use case and assessed feasibility through benchmarks across ML models of increasing size and comparing it with adjacent auditing methodologies as baselines.

Future work proceeds in two directions. First, we plan to validate the framework in real-world settings through industry case studies to assess its performance and scalability. Second, we will extend its capabilities by implementing auditing workflows for other trustworthy AI properties, such as verifying dataset fairness or model robustness against adversarial attacks.

\section*{Data Availability and Replication Package}
The source code of the framework, benchmarks and experiments, is available at \url{https://doi.org/10.5281/zenodo.20082491}.

\bibliographystyle{IEEEtran}
\bibliography{bibliography}

\end{document}